\documentclass[aps,prr,twocolumn,superscriptaddress]{revtex4-2}
 
\usepackage{amsmath}
\usepackage{amssymb}
\usepackage{mathtools}
\usepackage{amsthm}
\usepackage{tablefootnote}
\usepackage{float}
\usepackage{placeins}

\usepackage[linesnumbered,lined,boxed,ruled,commentsnumbered]{algorithm2e}
\usepackage{orcidlink}

\allowdisplaybreaks

\theoremstyle{plain}

\theoremstyle{definition}

\theoremstyle{remark}

\usepackage[textsize=tiny]{todonotes}

\usepackage[braket,qm]{qcircuit}
\usepackage{nkj}
\usepackage[utf8]{inputenc}
\usepackage[T1]{fontenc}  
\usepackage{xcolor}

\usepackage[textsize=tiny]{todonotes}

\usepackage{graphicx}
\usepackage{dcolumn}
\usepackage{bm}

\usepackage{hyperref}
\usepackage[nameinlink]{cleveref}

\begin{document}
\preprint{APS/123-QED}

\title{Bias Analysis and Regularization of Sequential Minimal Optimization \\ 
in Variational Quantum Eigensolvers}

\author{Samuele Pedrielli\orcidlink{0009-0005-5973-2235}}
\affiliation{Berlin Institute for the Foundations of Learning and Data BIFOLD, Berlin, Germany}
\affiliation{Technische Universität Berlin, Berlin, Germany}
\affiliation{Padova University, Padova, Italy}

\author{Frederik Stalschus\orcidlink{0009-0007-9196-0814}}
\affiliation{Berlin Institute for the Foundations of Learning and Data BIFOLD, Berlin, Germany}
\affiliation{Technische Universität Berlin, Berlin, Germany}

\author{Stefan Kühn\orcidlink{0000-0001-7693-350X}}
\affiliation{Deutsches Elektronen-Synchrotron DESY, Zeuthen, Germany}

\author{Karl Jansen\orcidlink{0000-0002-1574-7591}}
\affiliation{Deutsches Elektronen-Synchrotron DESY, Zeuthen, Germany}
\affiliation{Computation-Based Science and Technology Research Center, The Cyprus Institute, Nicosia, Cyprus}

\author{Kim A. Nicoli\orcidlink{0000-0001-5933-1822}}
\affiliation{Oldendorff Carriers GmbH $\&$ Co. KG, Lübeck, Germany}

\author{Shinichi Nakajima\orcidlink{0000-0003-3970-4569}}
\affiliation{Berlin Institute for the Foundations of Learning and Data BIFOLD, Berlin, Germany}
\affiliation{Technische Universität Berlin, Berlin, Germany}
\affiliation{RIKEN AIP, Tokyo, Japan}

\date{\today}

\begin{abstract}
The Nakanishi–Fujii–Todo (NFT) algorithm, also known as \texttt{Rotosolve}, implements Sequential Minimal Optimization for Variational Quantum Eigensolvers (SMO-VQE) by exploiting the trigonometric dependence of the energy on individual circuit parameters. This enables analytical one-dimensional minimization using only a few---typically two---energy evaluations, but introduces bias in the estimated energy. Although performing additional measurements every few tens of iterations can mitigate bias accumulation, we find that such corrections often degrade optimization performance. In this paper, we analyze the origin and accumulation of bias during the SMO-VQE process. Specifically, we show that the bias can be accurately estimated without additional measurements. Furthermore, we find that bias correction destabilizes optimization along directions with small curvature, whereas the original biased estimator implicitly acts as a regularizer. Based on these insights, we propose a simple regularization method that implements error accumulation while maintaining unbiased energy estimation. The resulting algorithm consistently improves performance across different system sizes, circuit depths, target Hamiltonians, and measurement shots, with minimal hyperparameter tuning.
\end{abstract}

\maketitle


\section{\label{sec:introduction}Introduction}

A longstanding challenge in quantum computing is understanding how currently available Noisy Intermediate-Scale Quantum (NISQ) devices can be leveraged to outperform classical machines on tasks of practical relevance. Prominent examples include the simulation of quantum many-body dynamics~\cite{Randall2021,Kim2023,Shinjo2026}
, lattice gauge theories~\cite{Farrell2024a,Farrell2024,Ciavarella2024,Alexandrou2025,meth_simulating_2025,Chai2025a,Chai2025,Gonzalez-Cuadra2025,Cochran2025,Crippa2026,Xu2026,Joshi2026}, computation of molecular ground-states~\cite{Peruzzo_2014,Kandala2017,GoogleAI2020,Robledo-Moreno2025}, and more general applications such as solving linear systems of equations~\cite{huang2019neartermquantumalgorithmslinear} or approximately solving combinatorial optimization problems~\cite{farhi_quantum_2014,Abbas2024}.

Among the approaches proposed to achieve a quantum advantage in the NISQ era, Variational Quantum Algorithms (VQAs) \cite{Cerezo_2021} have emerged as particularly promising. Their appeal lies in the combination of quantum state preparation with classical optimization, enabling the use of flexible, learning-based strategies tailored to near-term hardware constraints. In VQAs, a parametrized quantum circuit prepares a trial state, and a classical optimizer iteratively updates the circuit parameters to minimize a cost function estimated from quantum measurements.

A paradigmatic example is the Variational Quantum Eigensolver (VQE) \cite{Peruzzo_2014}, where the cost function is the expectation value of a Hamiltonian whose ground-state encodes the solution to the problem of interest. The performance of VQE depends on many design choices, including the initial state \cite{Grant_2019, anschuetz_quantum_2022}, the circuit ansatz \cite{Kandala_2017, romero_2018, Grimsley_2019}, and the classical optimization routine \cite{Bonet_Monroig_2023}. This work focuses on classical optimization and, in particular on the Nakanishi-Fujii-Todo (NFT) algorithm \cite{nft}, also known as \texttt{Rotosolve} \cite{Ostaszewski2021structure}, which implements Sequential Minimal Optimization (SMO) \cite{platt1998sequential} within the VQE framework.

The key insight behind SMO-VQE is that, for many commonly used parametrized gates, the energy expectation value depends sinusoidally on the individual circuit parameters. In practice, this functional form is not accessed directly, but inferred from noisy energy measurements. As a result, the energy landscape along a single parameter direction can be reconstructed by measuring the energy at few instances of circuit parameters, allowing the location of the one-dimensional minimum to be obtained analytically from the fitted sinusoidal model. SMO-VQE benefits from favorable theoretical \cite{Bertsekas01031997} and empirical convergence properties that position it as a solid alternative to gradient-based methods \cite{Singh_2023}. The algorithm has been successfully applied to a variety of problems, including the study of $\mathrm{SU}(N)$ fermions \cite{Consiglio_2022} and molecular correlation functions \cite{PhysRevA.104.032405}.

A crucial practical optimization introduced in the original SMO-VQE works exploits the fact that the minimum obtained along one direction lies in the one-dimensional subspace optimized in the subsequent step. This means that the obtained minimum can be reused as one of the energy evaluations required for the next one-dimensional fit. This significantly reduces the total number of measurements, thus reducing the total computational cost on a quantum device, which is a major advantage in the NISQ regime.

However, this efficiency gain comes at a cost. The reused minimum is an estimator of the minimum obtained from noisy measurements. As the algorithm progresses, statistical errors associated with these estimated minima accumulate and propagate through subsequent optimization steps. Although heuristic strategies for correcting accumulated statistical errors, such as periodically measuring the reused minimum, are commonly employed to mitigate this effect, the origin and accumulation of statistical errors has so far lacked a mathematical characterization.

In this work, we analyze the SMO-VQE algorithm from a Bayesian inference perspective, showing that the estimation of the minimum at each iteration introduces a bias directly linked to the statistical uncertainty of the measurements. We demonstrate that this bias leads to a progressive underestimation of the energy throughout optimization,
and that it can be corrected with an analytically-derived bias estimator.

Crucially, we also argue that removing this bias is not unconditionally beneficial. While bias correction improves the statistical consistency of the energy estimates, it simultaneously enhances the effect of observation noise, destabilizing the optimization process.
This is especially problematic along one-dimensional subspaces with small curvature, where a weak signal-to-noise ratio is synonym of poor quality updates. We find that optimization convergence \textit{does degrade when the energy estimator is unbiased}---either through additional measurements or bias correction. This is because biased estimators can act as an implicit regularization, stabilizing the optimization process.

Motivated by this insight, we develop a numerically tractable method to compensate for bias accumulation, and propose a simple regularization scheme that intentionally lowers the estimated energy when the one-dimensional fit is performed. Since the original uncontrolled bias is not necessarily optimal, deliberately allowing for a controlled amount of bias can improve optimization performance. 

We demonstrate that our proposed algorithm, using a single fixed hyperparameter, consistently improves upon the original NFT across a wide range of settings, including the number of qubits, the number of quantum circuit layers, target Hamiltonians, and the number of measurement shots, while retaining an unbiased energy estimate.

The remainder of this paper is structured as follows. In \Cref{sec:background}, we introduce the notation and review the fundamentals of VQE and SMO-VQE. \Cref{sec:biasinnft} introduces the Bayesian model and a detailed analysis of bias accumulation. In \Cref{sec:methods}, we present our correction and regularization methods. Finally, \Cref{sec:experiments} provides comprehensive experimental validations and results.

In the Appendixes we report the full analytical derivation of the bias expression in detail, together with its extension to more general circuits, a scaling analysis of our regularization method and its application to various spin Hamiltonians.  

\section{\label{sec:background}Background}
\subsection{\label{sec:nft} Variational Quantum Eigensolver}
VQE is a hybrid quantum–classical algorithm designed to approximate the ground-state of a Hamiltonian operator $H$. In VQE, a parametrized quantum circuit (ansatz) prepares a trial quantum state from a fixed initial state \footnote{More generally, the circuit structure itself may be adapted during optimization, as in \texttt{Rotoselect} \cite{Ostaszewski2021structure}.}. The circuit parameters are then optimized by a classical routine to minimize the expectation value of $H$.

Throughout this work, we consider circuits composed of fixed entangling gates (e.g. controlled-$X$) and independently parametrized single-qubit rotations of the form $U_d(\theta_d) = \exp\!\left(- i \frac{\theta_d}{2} O_d \right)$,
where the generators $O_d$ are Hermitian and unitary operators satisfying
$O_d^2 = \mathbf I$.

Each parameter $\theta_d$, with $d\in\{1,\dots,\mathcal D\}$, where $\mathcal{D}$ is the dimensionality of parameter space, parametrizes at most one gate \footnote{A generalization in which a parameter may control multiple gates is discussed in \Cref{appendix:vd}.}. The full circuit therefore defines a parametrized unitary operator $\mathcal G(\boldsymbol{\theta})$, which maps an initial state $\rho_{\mathrm{in}}$ to $\rho(\boldsymbol{\theta})=\mathcal G(\boldsymbol{\theta})\rho_{\mathrm{in}}\,\mathcal G^\dagger(\boldsymbol{\theta})$.
Since each parameter is an angle, the parameter space naturally forms a $\mathcal D$-dimensional torus $\boldsymbol{\theta}\in\mathbb T^{\mathcal D}$.

The VQE objective is therefore
\begin{align}
\boldsymbol{\theta}^{*}
=
\arg\min_{\boldsymbol{\theta}\in\mathbb T^{\mathcal D}}
\langle H\rangle(\boldsymbol{\theta}),
\label{eq:vqeobjective}
\end{align}
where
$\langle H\rangle(\boldsymbol{\theta})=\mathrm{Tr}\!\left[H\,\rho(\boldsymbol{\theta})\right]$.

The optimal parameters $\boldsymbol{\theta}^*$ correspond to the circuit that prepares a state approximating the ground-state $\rho_{\mathrm{GS}}$ of $H$, up to the expressive limitations of the chosen ansatz.

\subsubsection*{Measurement noise}

A defining feature of the optimization problem in \Cref{eq:vqeobjective} is that the objective function cannot be evaluated exactly. Instead, expectation values are estimated from quantum measurements and are therefore affected by statistical noise.
We distinguish two sources of noise:

\paragraph{Shot variance.}
This is the statistical uncertainty resulting from a finite number of shots (measurements), denoted by $N_{\mathrm{shots}}$. The corresponding error $\varepsilon_\mathrm{shots}$ can be approximated as Gaussian-distributed with variance scaling as $\sigma^2 \propto \frac{1}{N_{\mathrm{shots}}}$.
\paragraph{Hardware noise.}
This term encompasses errors arising from imperfect quantum hardware, such as gate infidelities and decoherence.

In this work we neglect hardware noise and focus exclusively on shot variance. This assumption is common in studies of optimization algorithms for VQEs \cite{nft,Ostaszewski2021structure,anders2025adaptiveobservationcostcontrol,pedrielli2025bayesianparametershiftrule} and remains relevant even in the fault-tolerant regime, where expectation values must still be estimated from finite sampling.
Under this assumption, each evaluation of the cost function yields an observation
\begin{align}
f(\boldsymbol{\theta})
=
\langle H\rangle(\boldsymbol{\theta})
+
\varepsilon_{\mathrm{shot}}.
\label{eq:measurement}
\end{align}

Because quantum measurements are costly, the computational cost of VQE optimization is typically quantified by the cumulative number of measurements performed.
Reducing the number of measurements required to obtain a good approximation of $\boldsymbol{\theta}^*$ is therefore a central goal of algorithmic research.


\subsection{\label{sec:nft}Sequential Minimal Optimization for Variational Quantum Eigensolvers}

A key structural feature of circuits composed of fixed entanglers and parametrized single-qubit rotations is that the energy landscape along any single parameter direction has an exact trigonometric form \cite{nft,Ostaszewski2021structure}.
Specifically, let
\begin{align}
\boldsymbol{\theta}_d
:=
(\theta_1,\dots,\theta_{d-1},\theta_d,\theta_{d+1},\dots,\theta_{\mathcal D})
\label{eq:theta-varies-d}
\end{align}
denote the parameter vector in which only the $d$-th component varies  while all other components remain fixed. In this one-dimensional subspace, the energy takes the form
\begin{align}
\langle H\rangle(\boldsymbol{\theta}_d)
=
b_{1,d}
+
\sqrt{2}
\big(
b_{2,d}\cos\theta_d
+
b_{3,d}\sin\theta_d
\big),
\label{eq:energyfunctionalform}
\end{align}
where the coefficients $\boldsymbol b_d=(b_{1,d},b_{2,d},b_{3,d})$ depend implicitly on the fixed parameters $\{\theta_i\}_{i\neq d}$.
The minimum of \Cref{eq:energyfunctionalform} occurs at
\begin{align}
\theta_d^{\mathrm{min}}(\boldsymbol b_d)
=
\operatorname{atan2}(b_{3,d},b_{2,d})+\pi.
\label{eq:minimizer}
\end{align}

Because the model contains three unknown coefficients, three energy evaluations are sufficient to determine $\boldsymbol b_d$ and compute the corresponding minimizer. In the noiseless limit $N_{\mathrm{shots}}\to\infty$ the exact coefficients $\boldsymbol b_d^*$ and minimizer
$\theta_d^*=\theta_d^{\mathrm{min}}(\boldsymbol b_d^*)$
are obtained. With finite sampling one instead obtains estimates $\hat{\boldsymbol b}_d$, from which the updated parameter vector
\begin{align*}
\hat{\boldsymbol{\theta}}_d
:=
(\theta_1,\dots,\theta_{d-1},
\theta_d^{\mathrm{min}}(\hat{\boldsymbol b}_d),
\theta_{d+1},\dots,\theta_{\mathcal D})
\end{align*}
and the corresponding estimated energy
$\hat f_d(\hat{\theta}_d)
:=
\langle H\rangle(\hat{\boldsymbol{\theta}}_d)$
are obtained.

We stress the distinction between 
\begin{itemize}
    \item \textbf{measured energies} $f(\boldsymbol{\theta})$, obtained from \Cref{eq:measurement} with a finite number of shots,
    \item \textbf{estimated energies} $\hat f(\theta)$, derived from \Cref{eq:energyfunctionalform} using estimated coefficients $\hat{\boldsymbol{b}}$, 
    \item \textbf{true energies} $f^*(\theta)$, derived from \Cref{eq:energyfunctionalform} using the true coefficients $\boldsymbol{b}^*$, or, equivalently, obtained from \Cref{eq:measurement} in the infinite shots limit.
\end{itemize}

The structure of \Cref{eq:energyfunctionalform} can be exploited when performing SMO-VQE to reduce the number of required measurements. Since the optimization proceeds sequentially across parameter directions, the minimum obtained along one axis can be reused as one of the observations required to fit the energy landscape along the next axis. Consequently, only two new energy measurements are required per iteration instead of three.

\subsubsection{\label{sec:SMO-VQE}The SMO-VQE routine}

Given an initial parameter vector $ \hat{\boldsymbol{\theta}}_0$ and its measured energy $f(\hat{\boldsymbol{\theta}}_0)$, the SMO-VQE algorithm iteratively minimizes the energy along successive parameter directions.
More specifically, at each iteration focusing on direction $d$, the algorithm proceeds as follows:

\begin{enumerate}

\item Measure the energy at two shifted points
$\boldsymbol{\theta}_d^{\pm\alpha}=\hat{\boldsymbol{\theta}}_{(d-1)\,\mathrm{mod}\,\mathcal{D}}\pm\alpha\,\mathbf e_d$ in direction $\mathbf e_d$, obtaining $f_d^{\pm\alpha}(\boldsymbol{\theta}_d^{\pm\alpha})=\langle H\rangle(\boldsymbol{\theta}_d^{\pm\alpha})+\varepsilon_{\mathrm{shot}}^{\pm\alpha}$.

\item Using the observations $f_d^{\pm\alpha}$ together with the reused estimate $\hat f_{d-1}$ (total of three points), fit the model in \Cref{eq:energyfunctionalform} to obtain $\hat{\boldsymbol b}_d$.

\item Compute the minimizer $\hat\theta_d=\theta_d^{\mathrm{min}}(\hat{\boldsymbol b}_d)$ by \Cref{eq:minimizer} and update the parameter vector.

\item Continue with axis $(d+1)\bmod\mathcal D$ until convergence.

\end{enumerate}

While reusing measurements substantially reduces the number of required evaluations, it introduces a serious side effect. The reused value $\hat f_{d-1}$ is itself an \textit{estimator} obtained from noisy data. Consequently, statistical errors originating in the estimation of $\hat{\boldsymbol b}_{d-1}$ propagate into subsequent optimization steps.
This phenomenon was already noted in the original SMO-VQE works \cite{nft,Ostaszewski2021structure}, although it was not analyzed in detail. In particular, Ref. \cite{nft} proposed periodically re-measuring the minimum energy in order to remove error accumulation.

In the next section we analyze this phenomenon from a Bayesian inference perspective. This viewpoint clarifies the mechanism responsible for the accumulation of errors, and allows us to derive an approximate correction term analytically.

\section{\label{sec:biasinnft}Bias Accumulation in SMO-VQE}
\subsection{\label{sec:bayesianregression}Bayesian Linear Regression}

To analyze the statistical properties of the SMO-VQE update, we interpret the trigonometric energy model in \Cref{eq:energyfunctionalform} within a Bayesian linear regression framework \cite{bishop2006pattern,nicoli2024physicsinformedbayesianoptimizationvariational}.

The exact energy along a single parameter direction follows the trigonometric form in \Cref{eq:energyfunctionalform}, with exact coefficients $\boldsymbol{b}^*$. Since we can only estimate this energy from a finite number of circuit measurements, we treat the measured energy $f$ at parameter value $\boldsymbol{\theta}_d$ (as defined in \Cref{eq:theta-varies-d}) as a noisy observation of the exact value 
\begin{align*}
p(f|\boldsymbol{\theta}_d,\boldsymbol b)
=
\mathcal N_1(f;\boldsymbol b^{T}\boldsymbol\phi(\boldsymbol{\theta}_d),\sigma^2).
\end{align*}
Here $\boldsymbol b^{T}\boldsymbol\phi(\boldsymbol{\theta}_d)$ denotes the exact energy if $\boldsymbol b = \boldsymbol b^*$, and $\sigma^2$ indicates the shot variance of the averaged energy estimator, while $\mathcal{N}_g$ indicates the $g$-dimensional Gaussian distribution. In other words, we adopt a Gaussian to describe the distribution of finite-shot energy measurements.

Our regression model uses the first-order $L_2$-orthonormal Fourier basis
\begin{align*}
\boldsymbol\phi(\boldsymbol{\theta}_d)^T
=
\left[
1,\,
\sqrt{2}\cos\theta_d,\,
\sqrt{2}\sin\theta_d
\right],
\end{align*}
which reproduces the functional form of the one-dimensional energy landscape in \Cref{eq:energyfunctionalform}. A non-informative, i.e., flat, prior is assumed for the regression coefficients $\boldsymbol b$.

Consider three sampling points $\Theta=(\boldsymbol{\theta}_{d-1},\boldsymbol{\theta}_d^{+\alpha},\boldsymbol{\theta}_d^{-\alpha})$, with feature matrix $\boldsymbol{\Phi}=\big[\boldsymbol\phi(\boldsymbol{\theta}_d),\boldsymbol\phi(\boldsymbol{\theta}_d^{+\alpha}),\boldsymbol\phi(\boldsymbol{\theta}_d^{-\alpha})\big]$, and measurement vector $\boldsymbol f=(f_d,f_d^{+\alpha},f_d^{-\alpha})^T$.
Then, the posterior distribution of the regression coefficients is also Gaussian and given by
\begin{align*}
p(\boldsymbol b|\Theta,\boldsymbol f)
=
\mathcal N_3(\boldsymbol b;\hat{\boldsymbol b},\boldsymbol{\Sigma_b}),
\end{align*}
with $\boldsymbol{\Sigma_b}=(\boldsymbol{\Phi}\boldsymbol S^{-1}\boldsymbol{\Phi}^T)^{-1}$ and
$\hat{\boldsymbol b}=\boldsymbol{\Sigma}_b\boldsymbol{\Phi}\boldsymbol S^{-1}\boldsymbol f$.
The shot variance is assumed to be constant along the one-dimensional projection. This is the homoscedastic assumption encoded in $\boldsymbol S=\sigma^2\mathbf I_3$, where $I_d$ is the $d$-dimensional identity matrix.

Under the flat prior, it is known that the posterior mean follows
\begin{align}
    \hat{\boldsymbol b}
    \sim
    \mathcal N_3(\boldsymbol b^*,\boldsymbol{\Sigma_b}),
    \label{eq:postmean}
\end{align}
where $\boldsymbol b^*$ denotes the true coefficients.
The fluctuations of $\hat{\boldsymbol b}$ described by \Cref{eq:postmean} propagate to the estimated minimizer and therefore to the energy estimate produced by SMO.

\subsection{\label{sec:biasanalysis}Analysis of the Bias}
We now analyze how statistical fluctuations in the regression coefficients induce a systematic bias in the estimated minimum energy.
This analysis is carried along a single one-dimensional subspace of the energy landscape.

It is convenient to introduce the amplitude $R:=\sqrt{2 b_2^2 +2 b_3^2}$ of the sinusoidal component of the function  (\ref{eq:energyfunctionalform}). The true amplitude corresponding to the exact coefficients $\boldsymbol b^*$ is denoted by $R^*$, while the estimated amplitude by $\hat{R}$.

The regression coefficients $\hat{\boldsymbol{b}}$ obtained from the Bayesian linear regression model in \Cref{sec:bayesianregression} determine both the estimated minimizer $\hat\theta$ and the corresponding estimated energy $\hat f(\hat\theta)$.
We define the bias as the expected difference between the estimated and true energies evaluated at the estimated minimizer
\begin{align}
\overline{\Delta f}(\hat{\theta}):=-\mathbb{E}\left[\Delta f(\hat \theta)\right]=-\mathbb E\left[f^*(\hat\theta)-\hat f(\hat\theta)\right],
\label{eq:NFTBias}
\end{align}
where the expectation is taken over the distribution in \Cref{eq:postmean}.
As shown in \Cref{appendix:first-order}, this expression can be rewritten as
\begin{align*}
\overline{\Delta f}(\hat{\theta})=\mathbb E\left[\left(R^*+\hat R\right)\left(1-\cos\hat\delta\right)\right],
\end{align*}
where $\hat\delta = \hat\theta-\theta^*$ denotes the error in the estimated minimizer. Both $\hat R$ and $\hat\delta$ depend on fluctuations in the regression coefficients $\Delta \boldsymbol{b} := \hat{\boldsymbol b}-\boldsymbol b^*$.

For simplicity, we assume equidistant sampling points $\alpha_\pm = \pm 2 \pi/3$, which were proven to be optimal in terms of optimization accuracy \cite{endo_optimal_2023} (see \Cref{appendix:equidistant_shift} and \Cref{appendix:biasprop} for more details on the implications of this choice). With this choice, the posterior covariance  becomes 
$\boldsymbol {\Sigma_b} = \sigma_b^2 \boldsymbol I_3 $ for
\begin{align*}
    \sigma^2_b := \mathbb{V}[\hat b_2] = \mathbb{V}[\hat b_3] = \frac{\sigma^2}{3}.
\end{align*}
Up to the first-order in $\hat\delta$, the amplitude estimator is unbiased $\mathbb{E}[\hat{R}] = R^* + \mathcal{O}\left(\hat{\delta}^2\right)$, while the minimizer error is approximately Gaussian $\hat\delta \sim\mathcal{N}_1\left(0, \xi^{-2}\right)$ with
$\xi=R^*/\sigma_b$ indicating the signal-to-noise ratio.

Combining these results yields the following expression for the bias
\begin{align}
\overline{\Delta f}(\hat{\theta})=-2R^*\xi^{-2}+\mathcal O(\hat\delta^2)
\label{eq:BiasEstimateApprox}
\end{align}
(see \Cref{appendix:first-order} for detailed derivation).
Note that, using the definition of the signal-to-noise ratio, the leading contribution equals
\begin{align}
    \overline{\Delta f}^{(1)} (\hat{\theta}) = -2\sigma_b^2 / R^*,
    \label{eq:BiasEstimate}
\end{align}
thus, the bias increases with statistical noise and becomes more pronounced when the energy landscape is flat.

The generalization of this result to non-independently parametrized gates is derived in \Cref{appendix:vd}.

\Cref{eq:BiasEstimate} provides a practical estimator of the bias: since $\mathbb{E}[\hat R]=R^*+\mathcal{O}(\hat \delta^2)$, the unknown amplitude $R^*$ can be replaced by its estimate $\hat R$ without affecting the leading-order accuracy.
Note that the bias is always negative, and therefore, the plain SMO-VQE algorithm systematically underestimates the minimum energy. 

Because the estimated minimum energy $\hat f$ is reused when fitting the energy landscape along the next parameter direction, the bias accumulates through iterations.
If the estimate entering the next step carries a bias 
$\overline{\Delta f}(\hat{\theta}_{d-1})$, then the fitted coefficients are shifted in expectation as
\begin{align}
\begin{cases}
\mathbb{E}[\hat b_{1,\Delta}] = \mathbb{E}[\hat b_1] - \overline{\Delta f}(\hat{\theta}_{d-1}),\\
\mathbb{E}[\hat b_{2,\Delta}] = \mathbb{E}[\hat b_2] - \sqrt{2}\,\overline{\Delta f}(\hat{\theta}_{d-1}),\\
\mathbb{E}[\hat b_{3,\Delta}] = \mathbb{E}[\hat b_3],
\end{cases}
\label{eq:biasprop}
\end{align}
(see \Cref{appendix:biasprop}), which in turn induces a systematic shift of the subsequent energy estimation and so on. 
This phenomenon is what we refer to as bias accumulation.

\section{\label{sec:methods}Methods}
\subsection{\label{sec:biascorrection}Bias Correction}

Bias accumulation in SMO-VQE has been previously observed in the literature \cite{nft,Ostaszewski2021structure}, although its origin and implications have not been systematically analyzed. In \cite{nft}, this effect is mitigated through a heuristic \textit{stabilization} procedure, in which the energy at the estimated minimizer is re-measured every $N_p$ iterations \footnote{In \cite{nft}, the authors use $N_p=32$.}. This strategy periodically re-anchors the optimization trajectory to the true energy landscape at the cost of investing a fraction $\frac{1}{2\,N_p +1}$ of the total budget. For typical problem sizes, where $N_\mathrm{tot} \sim10^7$, this corresponds to a substantial number of additional measurements dedicated solely to bias mitigation.

In contrast, our approach uses the analytical bias estimate derived in \Cref{eq:BiasEstimate} to correct the energy at every iteration. Specifically, we define the corrected estimator as
\begin{align}
    \hat{f}^{c}(\hat{\theta}) = \hat f (\hat{\theta})- \overline{\Delta f}^{(1)}(\hat{\theta}).\,
\end{align}
This correction provides an unbiased estimate without requiring additional measurements. Moreover, since the correction is applied at every step, it prevents the accumulation of bias throughout the optimization, eliminating the need for periodic stabilization and its associated cost.


\subsection{\label{sec:biascorrection}Regularization}

Up to this point, we have characterized the origin and accumulation of bias in SMO-VQE. We now investigate its effect on the optimization dynamics and show that bias accumulation can act as a spontaneous regularization mechanism.

In standard implementations, this effect is typically mitigated through periodic re-measuring of the energy minimum (see \Cref{sec:nft}). However, we observe that suppressing the bias accumulation is not always beneficial. In particular, in the later stage of the optimization, allowing bias accumulation can improve convergence.

To understand this, consider the effect of a negative bias on the reused estimate $\hat{f}(\hat \theta_{d-1})$. Such bias effectively increases the curvature of the model around $\hat \theta _{d-1}$. In statistical terms, this corresponds to a posterior distribution for the next minimizer that is more concentrated around the current minimizer. Equivalently, bias accumulation increases the effective signal-to-noise ratio, since the estimated amplitude $\hat{R}$ grows while the noise level remains unchanged.

As a consequence, exploration is progressively discouraged as the optimization progresses, reducing large parameter updates. While this effect may hinder performance in the early stages of the optimization, where exploration is important, it stabilizes the optimization near convergence by preventing the algorithm from escaping a good minimum.

As shown in \Cref{sec:experiments}, allowing bias to accumulate leads to improved performance in contrast to both heuristic stabilization and explicit bias correction method.

Motivated by this observations, we propose to replace the uncontrolled bias accumulation with an explicit regularization scheme.
Specifically, at each iteration we perform the following steps:
\begin{enumerate}
    \item Remove the spontaneous bias from the reused estimate with our correction method
    \begin{align*}
        \hat f_{d-1}(\hat \theta_{d-1}) \to \hat f^{c}_{d-1}(\hat \theta_{d-1})
    \end{align*}
    \item Introduce a controlled regularization term
    \begin{align*}
        \hat f^{c}_{d-1}(\hat \theta_{d-1})\to \hat f^{r}_{d-1}(\hat \theta_{d-1}) :=\hat f^{c}_{d-1}(\hat \theta_{d-1})-r
    \end{align*}
    \item Use the regularized value together with the measurements $f^{\pm \alpha}_d$ to compute the minimizer $\hat\theta_d$ and the corresponding energy estimate $\hat{f}_d(\theta_d)$.
    \item Correct the bias introduced by the regularization in the final energy estimate.
\end{enumerate}

The choice of the regularization strength $r$ defines our regularization strategy. Guided by the considerations above, we require that regularization should be weak in the early stages---where exploration is essential---and progressively increase during convergence, without saturating too quickly to avoid preventing improvement.

We therefore define
\begin{align}
 r := \frac{e^\tau}{N_\mathrm{shots}/n_P}\sqrt{\frac{t}{n_q}} \left(1-e^{-\frac{2t}{T}}\right),
 \label{eq:regularization}
\end{align}
where $e$ denotes Euler's constant and its exponent $\tau$ is a hyperparameter, $N_\mathrm{shots}/n_P$ is the number of shots allocated to each Pauli operator that constitutes the Hamiltonian (see \Cref{sec:experiments}), $t$ is the current iteration index, $n_q$ is the number of qubits, and $T=\mathcal{D}\cdot N_\mathrm{sweeps}$ is the total number of optimization steps. 

The exponential factor suppresses regularization at early times, while the polynomial scaling gradually increases its effect during convergence.

The scaling behavior of this regularization strategy with respect to $n_q$, the number of qubits, $n_l$, that of layers, and $N_\mathrm{shots}/n_P$, the number of shots-per-Pauli-operator, is analyzed in \Cref{appendix:reg-scaling}.

\section{\label{sec:experiments}Experiments}
\subsection{\label{sec:experiments/setup}Setup}
We evaluate the performance of the proposed method on the one-dimensional quantum Transverse Field Ising Model ($\mathrm{TFIM}$) with open boundary conditions,
\begin{align}
    H_{\mathrm{TFIM}} = j \sum\limits_{i}^{n_q-1}\sigma^x_i\sigma^x_{i+1} + h\sum_{i=1}^{n_q}\sigma^z_i,
    \label{eq:TFI}
\end{align}
where $\sigma^p$, $p \in \{x,y,z\}$ denote the Pauli operators. We consider the ferromagnetic version close to the critical regime by setting the Hamiltonian's parameters to $j=h=-1$.

Regarding our circuit ansatz, we use the \texttt{EfficientSU(2)} ansatz implemented in \textsc{Qiskit} \cite{Qiskit} with $Q=5$ qubits and $L=3$ layers. This results in a total of $D = 2Q(L+1)=40$ variational parameters. The ansatz satisfies the assumptions required for SMO, and initial parameters $\boldsymbol{\theta}$ are sampled uniformly at random. As discussed in \Cref{sec:biasanalysis}, we use equidistant sampling points $\alpha_\pm = \pm2\pi/3$.
Expectation values are obtained using $N_\mathrm{shots}=900$ measurements per circuit evaluation. This corresponds to allocating $100$ shots to each Pauli operator that composes the Hamiltonian in \Cref{eq:TFI}. Furthermore, we adopt the following convergence criterion: we stop the algorithm after $N_\mathrm{sweeps}=200$, corresponding to a total of $T=\mathcal{D}\cdot N_\mathrm{sweep}$ parameter updates. All results are averaged over $N_\mathrm{seeds}=100$ random initializations, and we report mean values together with one standard deviation.

To assess performance, we consider two metrics. The first is the \textit{true energy}, defined as the energy at the current best parameter $\hat{\boldsymbol{\theta}}$ obtained by using an infinite-shot realization of $E(\hat{\boldsymbol{\theta}})$. 

The second is the fidelity $F$ between the circuit output state $\rho(\hat{\boldsymbol{\theta}})$ and the ground-state subspace $\rho_\mathrm{GS}$ obtained via exact diagonalization of $H$ \footnote{in the presence of degeneracies, the ground-state subspace is defined as the span of those eigenvectors that correspond to eigenvalues degenerate with the minimum eigenvalue (within a fixed numerical tolerance of $10^{-5})$.}. We report both metrics in terms of their deviation from the optimum:
\begin{align}
\Delta\mathrm{Energy} 
& 
= E(\hat{\boldsymbol{\theta}}) - E_{GS},
\label{eq:DeltaEnergy}\\
\Delta \mathrm{Fidelity}
&
=
1-
F ,
\label{eq:DeltaFidelity}
\end{align}
Finally, to highlight the effect of bias, we show the energy estimate produced by the algorithm $\hat{f}(\hat{\theta})$ compared to the true energy at the estimated minimizer $f^*(\hat\theta)$.

\subsection{\label{sec:experiments/results}Results}

The goal of our experimental study is twofold: first, to validate the bias estimation theory developed in \Cref{sec:biasanalysis}, and second, to assess the optimization performance of our regularization strategies.
Details of the experimental setup are provided in \cref{sec:experiments/setup}. 

\paragraph{Bias correction}

\begin{figure}[h]
    \centering
    \includegraphics[width=0.9\linewidth]{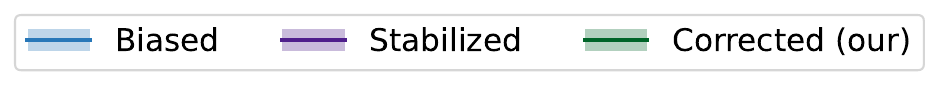}
    \includegraphics[width=0.9\linewidth]{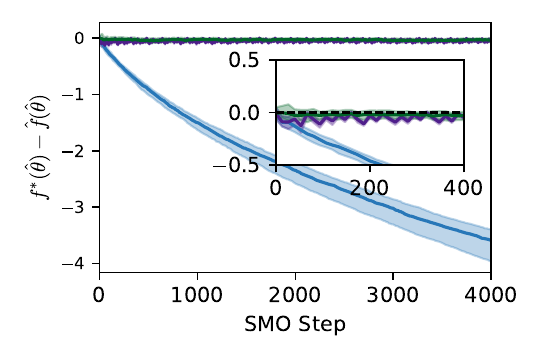}
    \caption{Estimation error of energy over SMO iterations by the biased, stabilized and our bias-corrected SMO-VQE algorithms.}
    \label{fig:biased-vs-stabilized-vs-corrected}
\end{figure}

In \Cref{fig:biased-vs-stabilized-vs-corrected}, we evaluate the accuracy of the estimated energy by the original biased SMO-VQE algorithm, the heuristic stabilization strategy of \ref{sec:nft}, and our analytic bias correction. 
Substantial error accumulation by the biased estimator is observed. Although the stabilized SMO-VQE improves the estimation error, periodic biases still remain (see the inset). In contrast, our corrected estimator accurately removes the bias. 

\paragraph{Bias regularization}

\begin{figure}[h]
    \centering
    \includegraphics[width=0.9\linewidth]{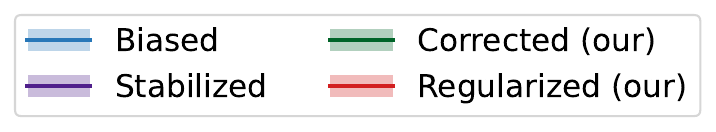}
    \includegraphics[width=0.9\linewidth]{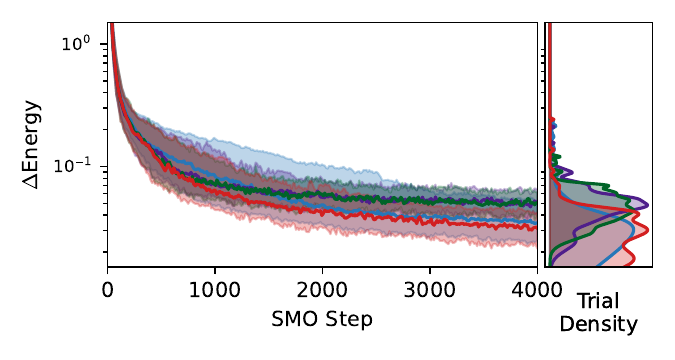}
    \includegraphics[width=0.9\linewidth]{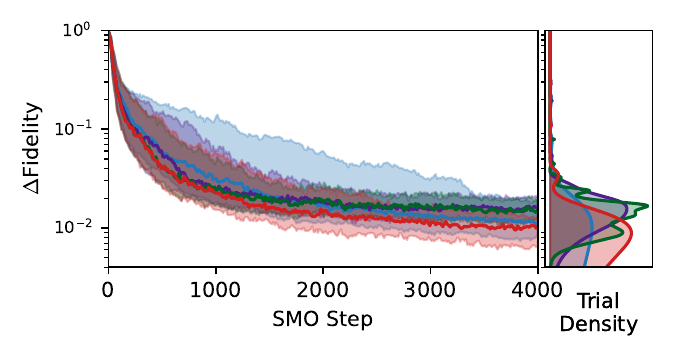}
    \caption{Performance comparison of the biased, corrected, and regularized SMO algorithms after $100$ sweeps, with $N_{\mathrm{shots}}=900$ per evaluation. 
    The spontaneous bias accumulation improves convergence relative to the stabilized and corrected algorithm, but is outperformed by the proposed regularization strategy.}
    \label{fig:5-3-900}
\end{figure}

\Cref{fig:5-3-900}
compares the optimization performance of our regularization strategy with baselines. As discussed in \Cref{sec:biascorrection},
removing the accumulated bias---which works as spontaneous regularization---either by stabilization or correction degrades the optimization performance.
However, the performance of the biased algorithm is itself suboptimal and our controlled regularization \eqref{eq:regularization}
consistently outperforms the original SMO-VQE, providing both unbiased energy estimation and better optimization performance across all phases of optimization.
We conducted evaluations for a variety of other circuit settings and Hamiltonians, and observed similar tendency in \Cref{appendix:reg-scaling}.



Importantly, we found that performance is not very sensitive to the hyperparameter $\tau$ that appears in \Cref{eq:regularization}. We set $\tau=2$ for all settings including the experiments in \Cref{appendix:reg-scaling}, and observe that our method always outperforms all baselines.
These results imply that our method is easily applicable to a wide range of problems without any parameter tuning or, if necessary, with a slight adjustment of $\tau$.


\section{\label{sec:conclusions}Conclusions}
In this work, we analyze the Nakanishi–Fujii–Todo (NFT) / \texttt{Rotosolve} algorithm for VQE, focusing on the reuse of energy estimates in sequential minimal optimization (SMO). While this reuse reduces the number of measurements required per update, it introduces a cumulative bias that systematically underestimates the energy. By formulating the one-dimensional subspace optimization step of the NFT algorithm as Bayesian linear regression problem, we identified this mechanism as a genuine bias accumulation effect.

In particular, we showed that the minimization step induces a systematic negative bias in the estimated minimum energy, which accumulates across iterations and eventually leads to unphysical energy estimates substantially below the true ground-state energy. In the high signal-to-noise regime and for the equidistant choice of shifts $\alpha_\pm=\pm 2\pi/3$, we derived an analytic approximation (\ref{eq:BiasEstimateApprox}) of the bias at each step, and characterized its propagation across coordinates via the coefficient update rule (\ref{eq:biasprop}). 
This provides a quantitative explanation for the empirical underestimation and the need for periodic stabilization observed in previous works.
Building on this analysis, we introduced a bias-corrected SMO-VQE update that subtracts an analytically-estimated bias term at each iteration. Unlike heuristic stabilization ---which requires additional measurements to periodically re-anchor the optimization trajectory--- our correction restores an unbiased energy estimate (to the leading order in $\hat\delta$) without increasing the quantum measurement cost. 

In numerical experiments on the Transverse Field Ising model using the \texttt{EfficientSU(2)} ansatz, this approach eliminates systematic underestimation without requiring extra measurements. Furthermore, we showed that uncontrolled bias accumulation can act as an implicit regularization mechanism: it suppresses late-stage exploration and can improve convergence, at the cost of producing biased energy estimates. 

Motivated by this insight, we proposed a bias-informed regularization strategy in which spontaneous bias is first removed, then replaced with a controlled negative offset prior to the update, which is also corrected after the update itself. This procedure preserves accurate energy estimates while retaining 
the beneficial effect of the bias. Empirically, the resulting regularized algorithm consistently outperforms both the unbiased and biased baselines across energy and fidelity metrics. 

These results highlight how a controlled bias can be leveraged as a useful design principle for improving performance in noisy quantum optimization.

\begin{acknowledgments}
This work was supported by the European Union’s HORIZON MSCA Doctoral Networks program project AQTIVATE (101072344) and the German Ministry for Education and Research (BMBF) under the grant BIFOLD25B

This work is funded by the European Union’s Horizon Europe Framework Programme (HORIZON) under the ERA Chair scheme with grant agreement no. 101087126.
This work is supported with funds from the Ministry of Science, Research and Culture of the State of Brandenburg within the Center for Quantum Technology and Applications (CQTA). 
\begin{center}
    \includegraphics[width = 0.3\textwidth]{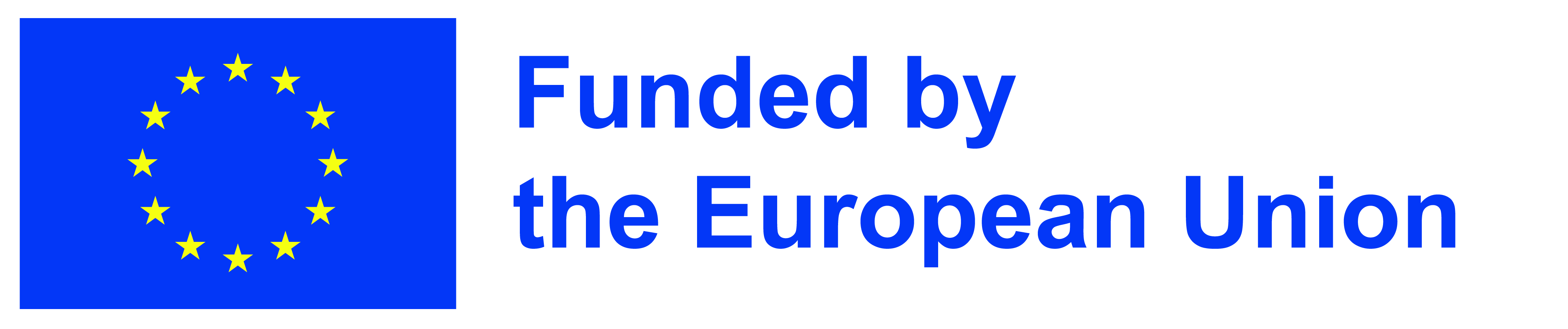}
    \includegraphics[width = 0.08\textwidth]{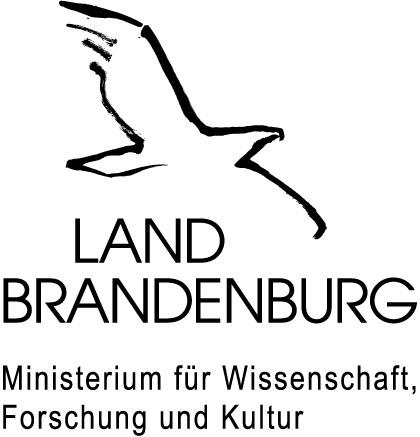}
\end{center}
\end{acknowledgments}

\appendix

\section{Equidistant measurements}
\label{appendix:equidistant_shift}

The configuration $\{\theta_1,\theta_-,\theta_+ \} = \{ \hat \theta, \hat \theta - 2\pi/3, \hat \theta + 2\pi/3\} $ was proven to be optimal in terms of optimization accuracy \cite{endo_optimal_2023}. This choice simplifies the structure of the covariance matrix $\boldsymbol{{\Sigma}_b}$ of the Bayesian posterior. In particular, the estimated coefficient components $\hat b_i$ become independent of each other and their variances $\sigma_{b}^{2}$ coincide.

Under the assumption that $\boldsymbol{S} = \sigma^{2}\boldsymbol{I}_{3}$, where $\sigma^2$ is the shot variance of the quantum measurement, the covariance matrix $\boldsymbol{\Sigma_b}$ becomes
\begin{align*}
\boldsymbol{\Sigma_{b}} 
& = 
\left( 
\boldsymbol{\Phi} 
\boldsymbol{S}^{-1}
\boldsymbol{\Phi}^T
\right)^{-1} \\
& = 
\sigma^2
\left(
\boldsymbol{\Phi}
\boldsymbol{\Phi}^T
\right)^{-1}\\
& =
\sigma^2
\left(
3\,
\boldsymbol{I}_{3}
\right)^{-1},
\end{align*}
where we used the fact that the rows of $\boldsymbol{\Phi}$ are orthogonal.
It follows that all off-diagonal elements of the covariance matrix are zero, while the diagonal is
\begin{align*}
    \mathbb{V}[\hat{b}_{i}]= \frac{\sigma^{2}}{3}:=\sigma^{2}_b, \hspace{2mm} \forall i \in \{1,2,3\}.
\end{align*}


\section{First-order expansion of the Bias}
\label{appendix:first-order}

In the following, we derive the expression for the bias $\overline{\Delta f}(\hat{\theta})$ as a function of the amplitude $R^*:=f^{*''}(\theta^*)$ of the energy projection and the signal-to-noise ratio $\xi$. In what follows, derivatives are indicated by $f^{'}(\theta):=\partial_\theta f(\theta)$.

Starting from \Cref{eq:NFTBias} and defining
\begin{align*}
    \Delta f(\theta) := f^*(\theta) - \hat f(\theta) \ ,
\end{align*}
we expand to first order in the minimizer displacement $\hat{\delta}$
\begin{align*}
    \Delta f(\hat\theta) = \Delta f (\theta^*) + \Delta f^{'}(\theta^*) \, \hat{\delta} + \mathcal{O}\left(\hat \delta ^2\right)
\end{align*}
and take expectations, getting
\begin{align}
    \mathbb{E}\left[\Delta f(\hat\theta)\right]= \mathbb{E}\left[ \Delta f^{'}(\theta^*) \, \hat \delta + \mathcal{O}\left(\hat \delta^2\right)\right]
    \label{eq:intermediate}
\end{align}
since the function is unbiased at any fixed (not estimated) input, i.e.,
\begin{align*}
    \mathbb{E}\left[\Delta f (\theta^*)\right]=0.
\end{align*}
Now we recall that, by definition,
\begin{align*}
    \hat{f}^{'}(\hat{\theta})=0 \implies f^{*'}(\hat \theta) - \Delta f^{'}(\hat\theta) = 0
\end{align*}
and expand both terms of this last equation around $\theta^*$
\begin{align*}
    f^{*'}(\hat \theta) &= f^{*'}(\theta^*) + R^* \, \hat\delta + \mathcal{O}\left(\hat\delta^2\right), \\
    \Delta f^{'}(\hat \theta) &= \Delta f^{'}(\theta^*) + \Delta f^{*''}(\theta^*) \, \hat\delta + \mathcal{O}\left(\hat\delta^2\right).
\end{align*}
From $f^{*'}(\theta^*)=0$ and $\Delta f^{''}(\theta^*) \, \hat\delta \sim \mathcal{O}\left( \hat\delta^2\right)$, we get 
\begin{align*}
    \hat\delta \approx \frac{\Delta f^{'}(\theta^*)}{R^*}.
\end{align*} 
Inserting this result in \Cref{eq:intermediate} and in the definition of the bias given by \Cref{eq:NFTBias} we obtain
\begin{align}
    \overline{\Delta f}(\hat{\theta}) &\approx -\mathbb{E}\left[\frac{1}{R^*}\left(\Delta f^{'}(\theta^*)\right)^2\right] \notag \\
    &= -\frac{1}{R^*}\mathbb{V}\left[\Delta f^{'}(\theta^*)\right], 
    \label{eq:variance}
\end{align}
which follows from $\mathbb{E}\left[\Delta f^{'}(\theta^*)\right]=0$. The variance is
\begin{align*}
    \mathbb{V}\left[\Delta f^{'}(\theta^*)\right]=2\left(\mathbb{V}[ \hat{b}_2]\cos^2\theta^* + \mathbb{V}[ \hat{b}_3]\sin^2\theta^*\right) = 2\sigma_b^2,
\end{align*}
where we used both the independence of the parameter estimates $\mathbb{E}\left[\hat b_2\right]\mathbb{E}\left[\hat b_3\right] = \mathbb{E}\left[\hat b_2\hat b_3\right]$ and the equality of their variances, as obtained in \Cref{appendix:equidistant_shift}.

The final expression for the bias is therefore
\begin{align*}
    \overline{\Delta f}(\hat{\theta}) \approx -\frac{2\sigma^2_b}{R^*}=-2R^*\xi^{-2}.
\end{align*}

\section{Generalization to non-independently parametrized gates}
\label{appendix:vd}
In the more general case where the parameter $\theta$ is shared among at most $N$ gates, the trigonometric form of the energy expectation is preserved, but higher order harmonics appear \cite{nft}
\begin{align*}
    f^*(\theta)=b_1+\sqrt{2}\sum\limits_{n=1}^{N}\left(b_n^{(c)}\cos\left(n\theta\right) + b_n^{(s)}\sin\left(n\theta\right)\right).
\end{align*}
Fitting $M=2N+1$ noisy energy observations at equidistant points gives the estimated coefficients $\hat{\boldsymbol{b}}$ and, consequently, the estimated function $\hat{f}$. 

As in the $N=1$ case, 
\begin{align*}
    \hat\delta \approx \frac{\Delta f^{'}(\theta^*)}{R^*},
\end{align*}

with the same conventions used in \Cref{appendix:first-order}.

Differentiating $f^*$ and $\hat f$, we get
\begin{align*}
    \Delta f^{'}(\theta^*) = \sum\limits_{n=1}^{N}\sqrt{2}n\left(-\Delta b_n^{(c)}\sin\left(n\theta^*\right) + \Delta b_n^{(s)}\cos\left(n\theta^*\right)\right).
\end{align*}

In order to compute the variance of this term, 
we recall that, for an odd number $M=2N+1$ of equidistant points $\{\theta_j\}_{j=0}^{M}$,
\begin{align*}    \sum\limits_{j=0}^{M-1}\cos^2(n\theta_j)=\sum\limits_{j=0}^{M-1}\sin^2(n\theta_j) = \frac{M}{2},
\end{align*}
which is a generalization of the result we used in \Cref{appendix:equidistant_shift}.
This implies that
\begin{align*}
    \mathbb{V}[\hat b^{(c)}_n]=\mathbb{V}[\hat b^{(s)}_n]=\frac{\sigma^2}{M}=\sigma^2_b.
\end{align*}

Using this fact, we obtain
\begin{align*}
    \mathbb{V}[\Delta f^{'}(\theta^*)] &= \sum\limits_{n=1}^{N}2n^2\left( \sin^2(n\theta^*) + \cos^2(n\theta^*)\right)\sigma^2_b \\
    &=2\sigma^2_b\sum\limits_{n=1}^{N}n^2,
\end{align*}
and the expression of the bias, given in \Cref{eq:variance}, 
 follows
\begin{align*}
    \overline{\Delta f}(\hat\theta) \approx 
    -\frac{2\sigma^2_b}{R^*}\sum\limits_{n=1}^{N}n^2.
\end{align*}

\section{Signal-to-noise regimes}
\label{a:snr}
As mentioned in \Cref{sec:biasanalysis}, our expression for the bias \Cref{eq:NFTBias} holds in the high signal-to-noise $\xi:=\frac{R^*}{\sigma^2}\gg 1$ regime. 

In this work, noise is assumed to be uniform along the one-dimensional energy projections. We approximate the variance along the subspace with the average of the two variances measured at $\boldsymbol{\theta}_d^{\pm2\pi/3}$ (see \Cref{sec:SMO-VQE}). This is an approximation since, in general, the variance is a function of the parameter values.

Although most parameter directions show a high signal-to-noise ratio, justifying our expansion, some directions of the chosen circuit ansatz have an energy curvature (signal) of the same magnitude as the noise, i.e. $\xi\sim1$. Along these flat directions, the variance of the estimated minimizer is of order $\sim \mathcal{O}\left(1\right)$. 

In principle, this could severely hinder optimization since the algorithm effectively chooses the minimizer at random along these directions. Interestingly, the spontaneous accumulating bias works as a natural remedy for this problem: by progressively underestimating the previous energy minimum, the resulting estimated curvature (which the algorithm then uses to estimate the next minimizer) grows, solving the flat direction issue.

This explains why the biased version of the algorithm performs better than its corrected counterpart in the converging phase and justifies our regularization approach as a controlled remedy for flat directions when using an unbiased estimator.

\section{Linear propagation of the bias}
\label{appendix:biasprop}
The expression of bias propagation given in \Cref{eq:biasprop} is also a consequence of the chosen sampling configuration $\{\hat \theta, \hat \theta - 2\pi/3, \hat \theta + 2\pi/3\}$, since with this choice the estimated coefficients are
\begin{align*}
\begin{cases}
    \mathbb{E}\left[\hat b_1\right] = \mathbb{E}\left[\hat f(\hat\theta_{d-1})\right] + f_{d}^{-2\pi/3}(\boldsymbol{\theta}_d^{-2\pi/3}) + f_{d}^{2\pi/3}(\boldsymbol{\theta}_d^{2\pi/3})\\
    \mathbb{E}\left[\hat b_2\right] =  \sqrt 2 \left( \mathbb{E}\left[\hat f(\hat\theta_{d-1})\right] - \frac{f_{d}^{-2\pi/3}(\boldsymbol{\theta}_d^{-2\pi/3}) + f_{d}^{2\pi/3}(\boldsymbol{\theta}_d^{2\pi/3})}{2}\right)\\
    \mathbb{E}\left[\hat b_3\right] = \sqrt{\frac{3}{2}}\left( f_{d}^{2\pi/3}(\boldsymbol{\theta}_d^{2\pi/3}) - f_{d}^{-2\pi/3}(\boldsymbol{\theta}_d^{-2\pi/3})\right),
\end{cases}
\end{align*}
where $f_d^{\pm2\pi/3}(\boldsymbol{\theta}_d^{\pm2\pi/3})$ follow the definition given in \Cref{sec:SMO-VQE}.

Having a biased estimate $\mathbb{E}\left[\hat f_{d-1}\right] = f^*_{d-1} - \overline{\Delta f}_{d-1}$ therefore results in the biased coefficients reported in \Cref{eq:biasprop}.

\section{\label{appendix:reg-scaling}Regularization Scaling}
In this section, we address the scaling of our regularization method to larger system sizes of the Transverse Field Ising Model described in \Cref{eq:TFI}.

We stick to the \texttt{EfficientSU(2)} ansatz for the parameterized circuits considered. The total number of parameters $\mathcal{D}$ in the circuit follows $\mathcal{D}=2\,n_q\left(n_l+1\right)$. We set $N_\mathrm{sweeps}=200$ to compare the algorithms deep in the converging phase.

\begin{figure}
    \center
    \includegraphics[width=1.\linewidth]{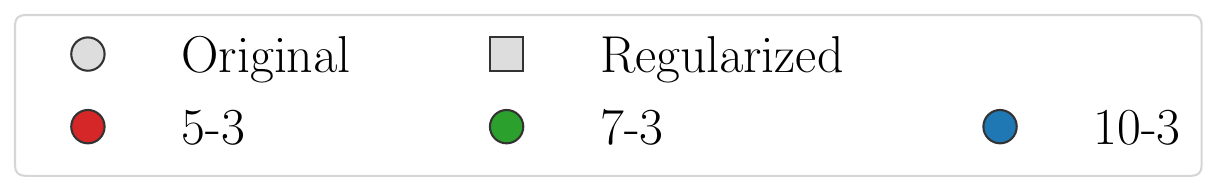}
    \includegraphics[width=1.\linewidth]{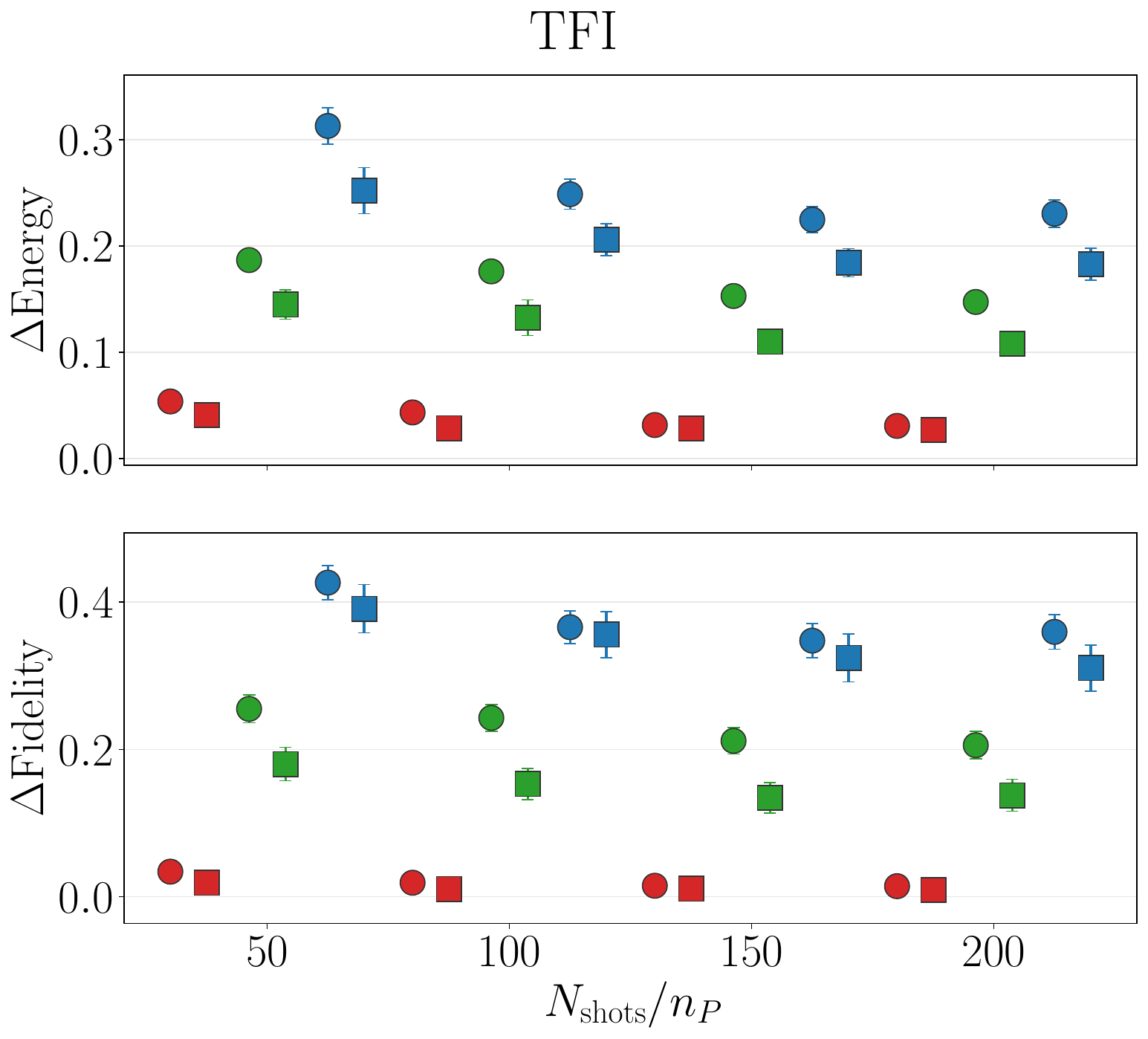}
    \caption{Comparison of the final $\Delta \mathrm{Energy}$ and $\Delta\mathrm{Fidelity}$ obtained with the original algorithm and our regularized method for $N_\mathrm{shots}/n_P \in \{50, 100, 150, 200\}$, $n_q\in\{5,7,10\}$ and $n_l=3$.}
    \label{fig:scaling-regularization-nq}
\end{figure}

\Cref{eq:regularization} expresses the dependency on $n_q$, the number of qubits,  and $N_\mathrm{shots}/n_P$, the shots-per-Pauli-operator.
Fig. \Cref{fig:scaling-regularization-nq} shows the final $\Delta\mathrm{Energy}$ and $\Delta\mathrm{Fidelity}$ obtained with our regularized algorithm (boxes) versus the original biased algorithm (circles) as a function of $N_\mathrm{shots}/n_P$ and $n_q$. 

The regularized algorithm not only consistently outperforms its original counterpart, but often performs better than the original algorithm when using fewer $N_\mathrm{shots}$. For instance, in the $7-3$ and $10-3$ settings, both regularized algorithms at $N_\mathrm{shots}/n_P=100$ perform better than their biased counterpart even when these use $150$ or even $200$ shots-per-Pauli-operator.

The reason behind this surprising improvement might lie in the effective increase in signal-to-noise ratio witnessed by the SMO-VQE update, especially along flat directions.

\begin{figure}
    \center
    \includegraphics[width=1.\linewidth]{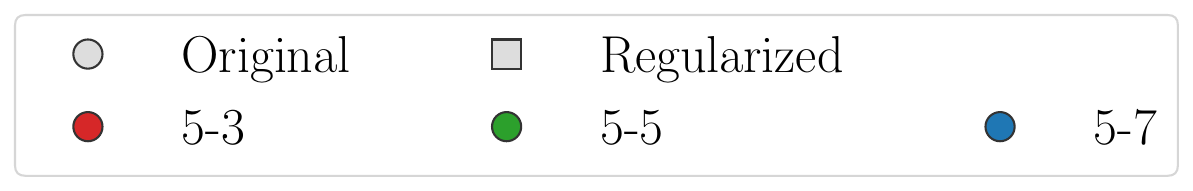}
    \includegraphics[width=1.\linewidth]{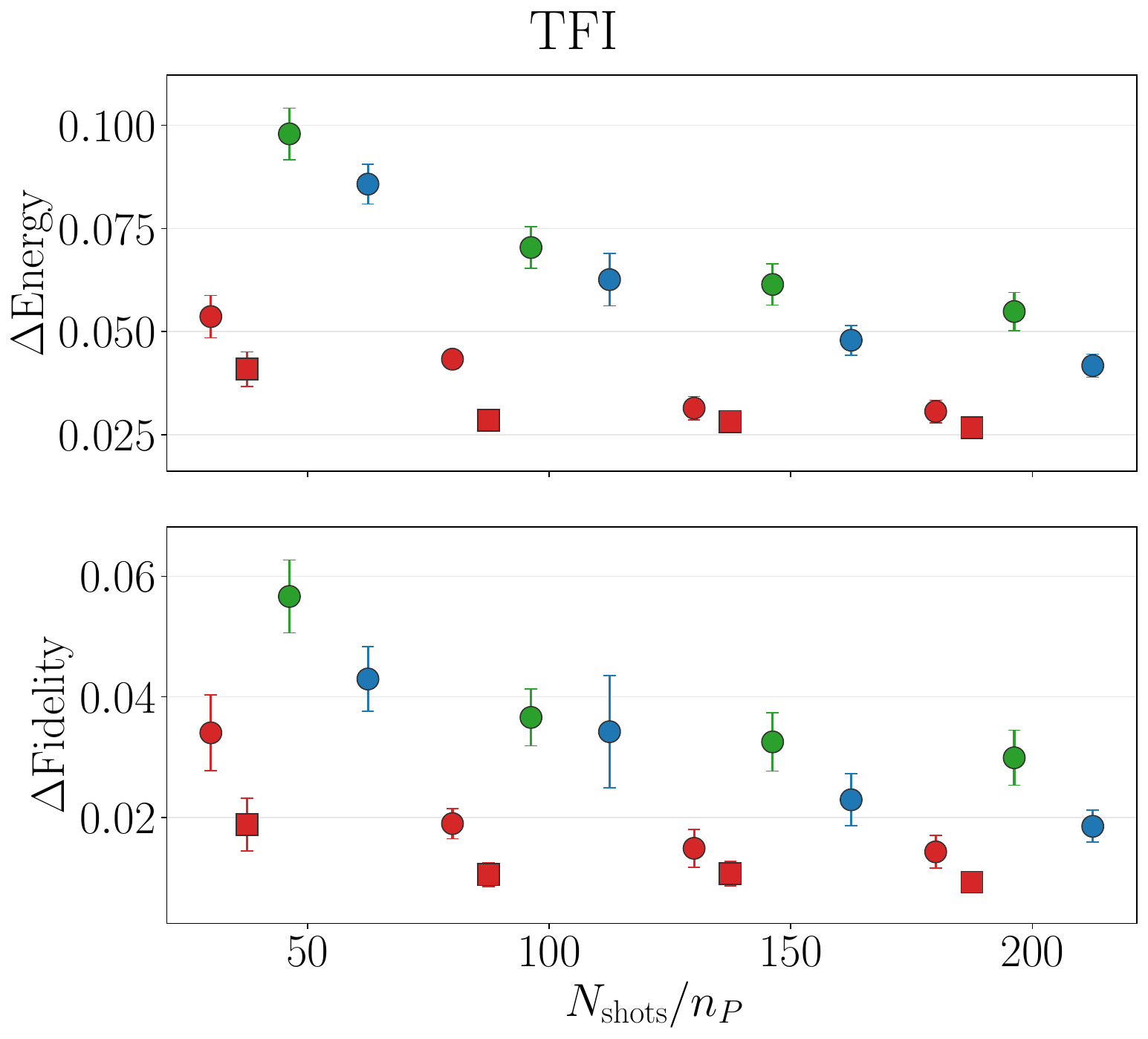}
    \caption{Comparison of final $\Delta \mathrm{Energy}$ and $\Delta\mathrm{Fidelity}$ obtained with the original algorithm and our regularized method 
    for $N_\mathrm{shots}/n_P  \in \{50, 100, 150, 200\}$, $n_q=5$, and $n_l\in\{3,5,7\}$.}
    \label{fig:scaling-regularization-nl}
\end{figure}

Fig. \Cref{fig:scaling-regularization-nl} shows the same quantities 
for different number of layers $n_l$. The regularized algorithm is reported only for the simplest case of $(n_q,n_l)=(5,3)$ since this is enough to outperform all other settings.

\section{\label{appendix:more-spin-hamiltonians}More Spin Hamiltonians}
In this section, we apply our correction and regularization methods to more complex spin Hamiltonians, demonstrating the generality of our results.  

In particular, we investigate the following spin chains for $(n_q,n_l)=(5,3)$.
\begin{itemize}
    \item $\mathrm{XX}$
    \begin{align*}
        H_{\mathrm{XX}} = 
        &\,j \left(
        \sum\limits_{i}^{n_q-1}\sigma^x_i\sigma^x_{i+1} +
        \sum\limits_{i}^{n_q-1}\sigma^y_i\sigma^y_{i+1}\right)
    \end{align*}
    for $j=-1$, which is an instance of the Luttinger Liquid phase. This model is known to be analytically solvable via a Jordan-Wigner mapping to free fermions.
    \item $\mathrm{XXZ}$ 
    \begin{align*}
        H_{\mathrm{XXZ}} = 
        &\,j \left(
        \sum\limits_{i}^{n_q-1}\sigma^x_i\sigma^x_{i+1} +
        \sum\limits_{i}^{n_q-1}\sigma^y_i\sigma^y_{i+1}\right) +
        \Delta \sum\limits_{i}^{n_q-1}\sigma^z_i\sigma^z_{i+1}
    \end{align*}
    for $j=2\Delta=-1$, which is again an instance of the Luttinger Liquid phase. While $\Delta=0$ (see XX model) implies a mapping to free fermions, here $\Delta\neq0$ results into a mapping to interacting fermions \cite{Giamarchi2004QuantumPI}.
    \item XXX or Isotropic Heisenberg Model
    \begin{align*}
        H_{\mathrm{Heis}} = 
        &\,j \left(
        \sum\limits_{i}^{n_q-1}\sigma^x_i\sigma^x_{i+1} +
        \sum\limits_{i}^{n_q-1}\sigma^y_i\sigma^y_{i+1} +
        \sum\limits_{i}^{n_q-1}\sigma^z_i\sigma^z_{i+1}\right) \, + \\
        &h \left(
        \sum_{i=1}^{n_q}\sigma^x_i +
        \sum_{i=1}^{n_q}\sigma^y_i +
        \sum_{i=1}^{n_q}\sigma^z_i
        \right)
    \end{align*}
    for $j=h=-1$, which corresponds to a fully polarized ferromagnet along the direction of the magnetic field $\mathbf{h}=(-1,-1,-1)$. 
\end{itemize}

\begin{figure}
    \center
    \includegraphics[width=1\linewidth]{figures/hamiltonians/scaling-regularization-legend.pdf}
    \includegraphics[width=1\linewidth]{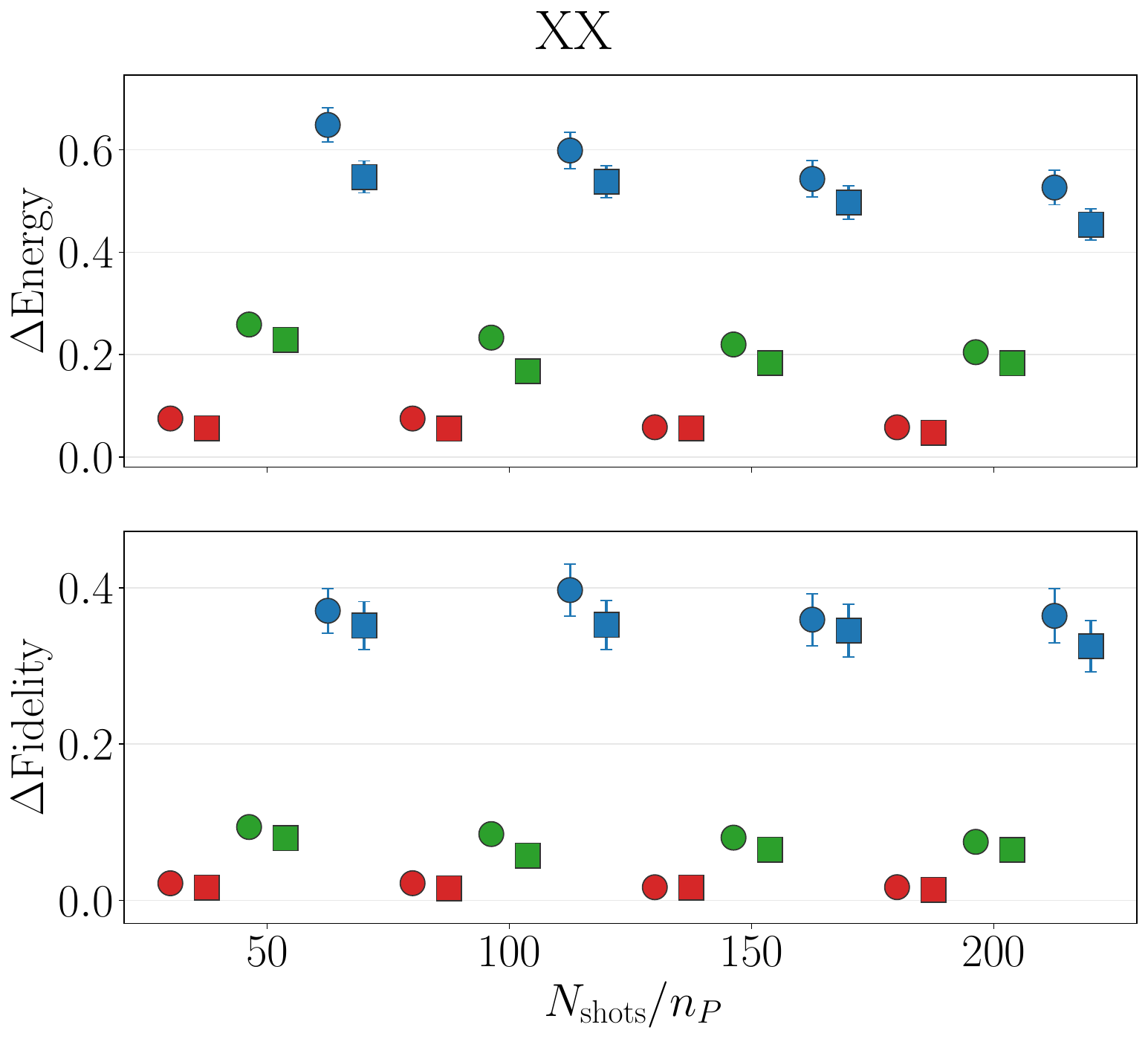}
    \caption{
     Comparison between the original algorithm and our regularization method on the XX model
for $N_\mathrm{shots}/n_P \in \{50, 100, 150, 200\}$, $n_q\in\{5,7,10\}$ and $n_l=3$.
    }
    \label{fig:scaling-regularization-xx-ll}
\end{figure}

\begin{figure}
    \center
    \includegraphics[width=1\linewidth]{figures/hamiltonians/scaling-regularization-legend.pdf}
    \includegraphics[width=1\linewidth]{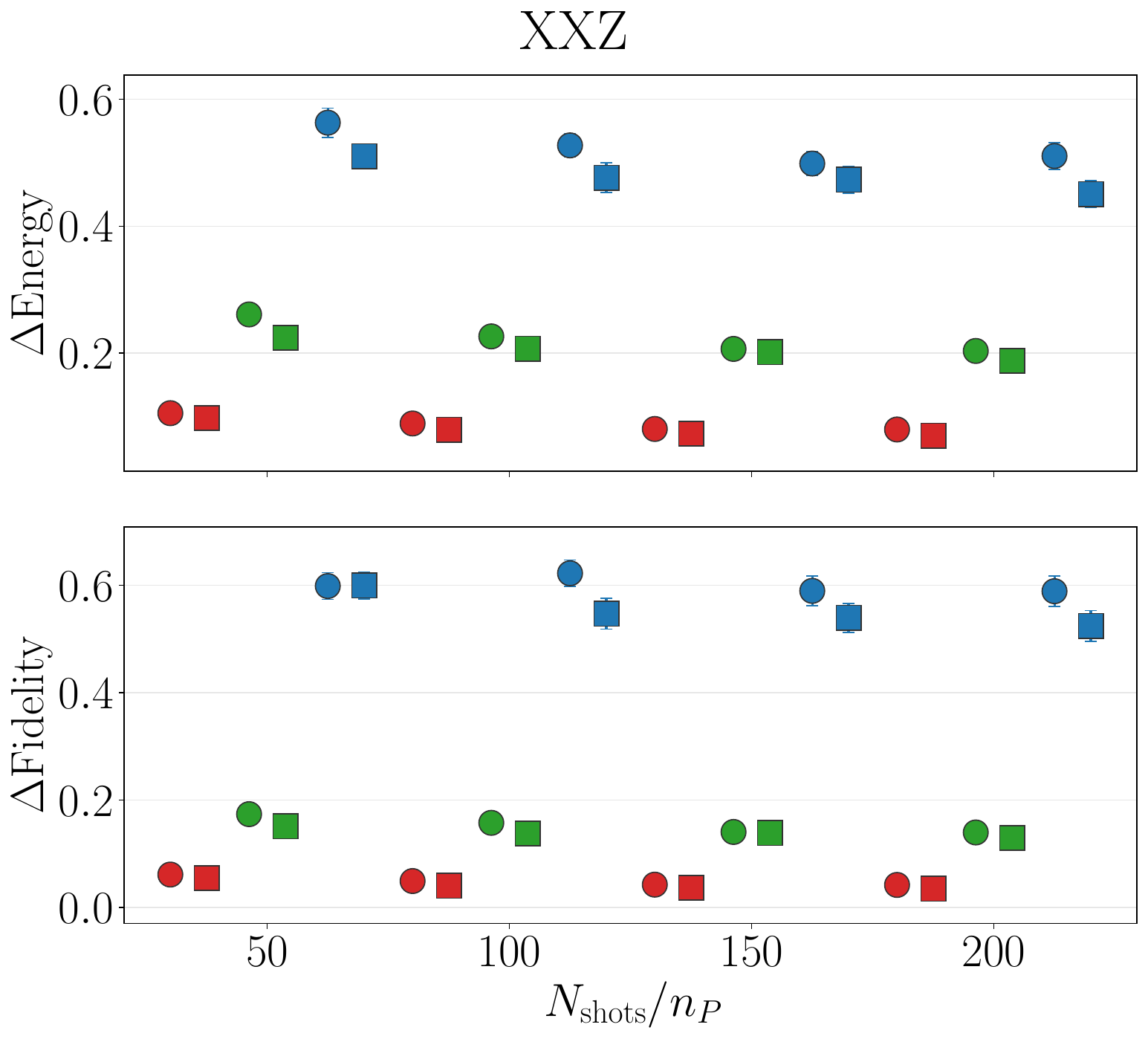}
    \caption{
    Comparison between the original algorithm and our regularization method on the XXZ model
for $N_\mathrm{shots}/n_P \in \{50, 100, 150, 200\}$, $n_q\in\{5,7,10\}$ and $n_l=3$.
    }
    \label{fig:scaling-regularization-xxz-ll}
\end{figure}

\begin{figure}
    \center
    \includegraphics[width=1\linewidth]{figures/hamiltonians/scaling-regularization-legend.pdf}
    \includegraphics[width=1\linewidth]{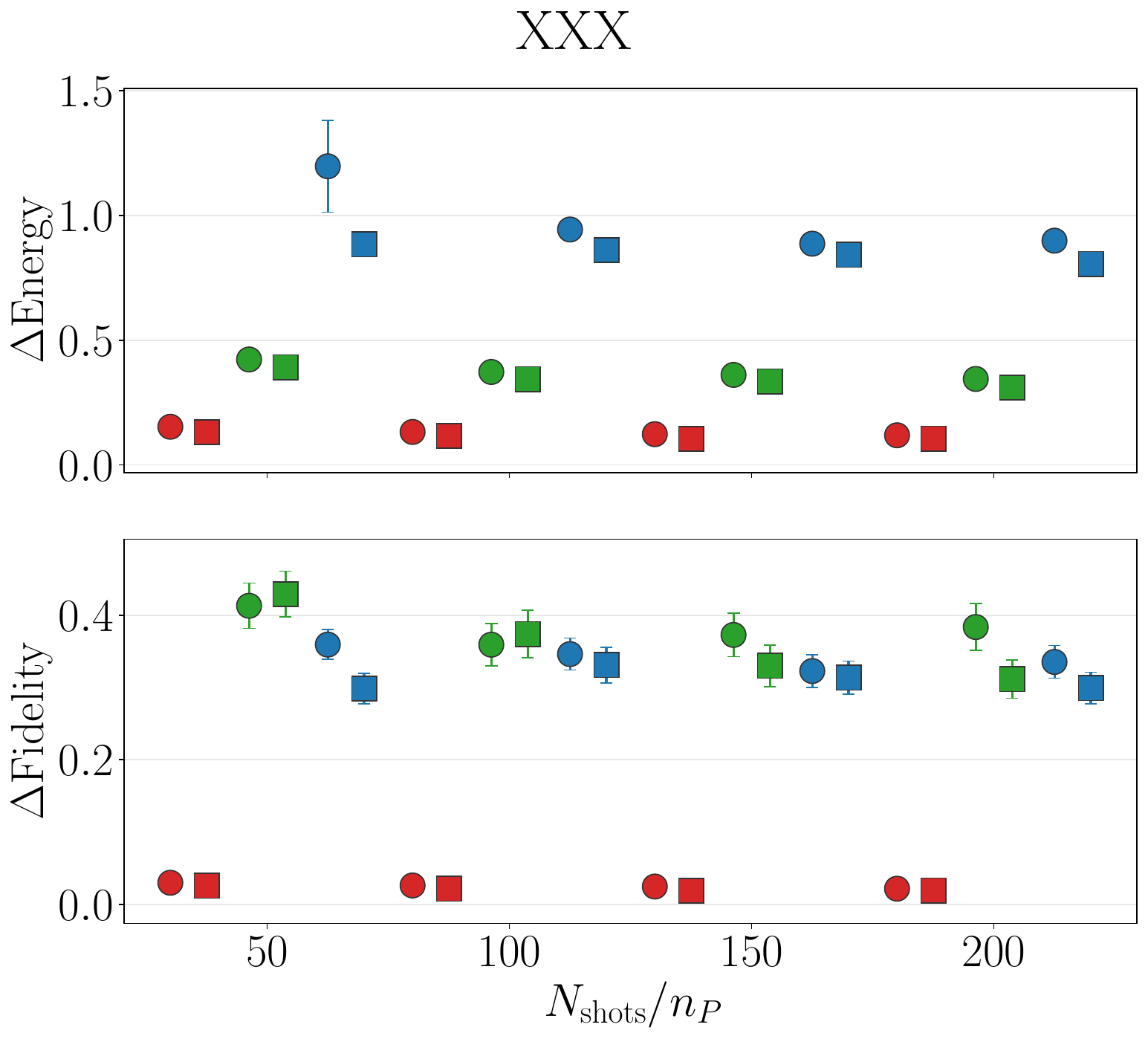}
    \caption{
    Comparison between the original algorithm and our regularization method on the XXX model
for $N_\mathrm{shots}/n_P \in \{50, 100, 150, 200\}$, $n_q\in\{5,7,10\}$ and $n_l=3$.
    }
    \label{fig:scaling-regularization-xxx-ll}
\end{figure}

In the spirit of comparing problems with similar noise levels, we set $N_\mathrm{shots}/n_P\in\{50,100,150,200\}$ for all Hamiltonians.
Figures \ref{fig:scaling-regularization-xx-ll}--\ref{fig:scaling-regularization-xxx-ll}
 show that our regularized algorithm consistently outperforms the original algorithm across all metrics and Hamiltonians.

\clearpage
\bibliography{biasbib}

\end{document}